\documentstyle[editedvolume]{crckapb}

\begin{opening}
\title{Mm Observations of IRAS Galaxies: Dust
Properties, Luminosity Functions and Contributions to the Sub-Mm
Background}

\author{P.ANDREANI \& A.FRANCESCHINI}
\institute{Dipartimento di Astronomia di Padova\\
vicolo dell'Osservatorio 5,
 I-35142 Padova, Italy.}

\end{opening}

\runningtitle{Mm Observations of IRAS Galaxies}

\begin{document}


\begin{abstract}
We have studied the FIR/{\it mm} spectrum of IR galaxies by combining IRAS
photometry with new {\it mm} data on a complete southern IRAS galaxy sample.
The observed spectra and a dust model emphasize a dicothomy in the galaxy
population: half of the objects with a lot of warm dust are
characterized by higher values of the bolometric (UV-FIR) luminosity, of the
dust-to-gas mass ratio, of the dust optical depths and extinction, 
while those dominated by cold ({\it cirrus}) dust show opposite
trends. From these data we derive the {\it mm} luminosity function of
galaxies and estimate their contribution to the sub-{\it mm} background (BKG).
\end{abstract}

\section{Introduction}

Observations of diffuse dust in galaxies impact
on some basic questions about their present structure and past history. In 
particular,
are we missing significant amounts of luminous matter because of the effects
of dust extinction? How much severe are the corresponding selection effects? 
 
Traditional approaches to investigate dust in galaxies rely on either 
{\it (a)} indirect estimates based on dust extinction effects in the optical 
and near-IR,
or {\it (b)} direct measurements of dust emission from FIR/mm observations.
Neither have provided conclusive results at the moment:
the former method is based on model-dependent and controversial 
assumptions about the optical-IR spectrum of various galactic components, the
latter still lacks a large enough 
statistical basis and suffers from the observational uncertainties 
of the sub-{\it mm}/{\it mm} data.
 
We have tackled the question of the dust content of spirals by observing
the 1.25 {\it mm} continuum emission from a complete sample of IRAS galaxies
with the SEST Telescope. The SEST telescope
was chosen as it provided the best compromise between detector
sensitivity and spatial resolution (see for more details
Franceschini \& Andreani 1995).
 
\section{The FIR/mm spectrum of galaxies: cold dust and extinction}
 
Our observations allowed us to directly estimate, for the first time,
the long-$\lambda$ spectrum of galaxies (see Figure 3 below) in a poorly 
explored spectral domain. 
We interpret the FIR/mm spectra exploiting a standard dust model. Two major
dust components are assumed to be present in the ISM: a cold "cirrus"
component and warm dust in star-forming regions. 
 
Sixteen galaxies, hereafter called {\it "cirrus"-dominated}, have a negligible
contribution of warm dust at 100 $\mu m$, while the other 
14 objects require high values of the warm dust 
fraction and their spectrum at $\lambda < 100\ \mu m$ is 
dominated by the starburst component. FIR/{\it mm} data and a reliable dust
model provide a straightforward way to classify galaxies according to their
star formation activity. 
The starburst component contributes only a couple of percent on average of the
total dust mass, while the cold component bears 97 \% of the total mass. 

\begin{figure}                
 \vspace{132pt}
\caption{Observed dust optical-depth $\tau_B$ versus overall 
extinction $A_B$. Open and filled squares refer to the 
inactive {\it cirrus}-dominated and to the starbursting objects,
respectively.
Position marked ($\odot$) corresponds to our Galaxy.
Predicted dependences for a screen, a slab, and a sandwich (zero dust
scale-height) model are shown for comparison (reprint from
Franceschini \& Andreani 1995).
}
\end{figure}

Figure 1 shows the extinction estimated from our data
versus the observed B-band optical depth. This
was found from the total dust mass divided by the galaxy
projected area and measures the amount of dust {\it available} to absorb 
the optical light. $A_B$ measures the overall 
{\it actual} effect of extinction and was
estimated from the logarithmic ratio of the bolometric optical-UV 
luminosity to the bolometric UV/FIR light.
 
Two galaxy classes are significantly
segregated over this plane: the inactive {\it "cirrus"-dominated}
objects being
confined to lower values of dust optical depth ($\tau_B < 2$) and low
extinction ($A_B \leq 1$). The active star-forming galaxies, on the
contrary, are spread over much larger values in both axes. 
Appreciable amounts of dust and  extinction seem
to characterize only the population of starbursting galaxies.
This may reflect a more exhaustive processing of the ISM through 
star-formation in the latter objects than for the average {\it inactive}
spiral.

\section{The {\it mm} luminosity function and contributions to the BKG} 
 
We find the mm emission for our sample galaxies to be very 
well correlated with 
the 60 and 100 $\mu m$ ones. The estimate of the mm 
local luminosity functions (LLF) is then straightforward, in the 
presence of a complete flux-limited sample.

\begin{figure}	
\vspace{130pt}
\caption{The local luminosity functions at 60 $\mu m$, 
compared with published data (panel [a]), and at 1330 $\mu m$
(panel [b]), estimated from our IRAS galaxy sample.
Open boxes in panel (b) are based on our new 60 $\mu m$ LLF,
whereas the filled ones are based of the Saunders's et al. LLF. 
The thick line 
is the 1300 $\mu m$ function transformed from the 60 $\mu m$ one through a 
constant $L_{1300}/L_{60}$ ratio.} 
\end{figure}

We report in Figure 2 the LLFs at 60 and 1300  $\mu m$
derived from our sample.
The good match of our 60 $\mu m$ LLF with published data (Saunders et 
al. 1990) emphasizes the completeness and reliability of the sample.
The 1300 $\mu m$ LLFs were derived from those at 60 
$\mu m$ by means of the 
bivariate IR/mm luminosity distribution (open and filled squares), and from
a simple scaling of the 60 $\mu m$ one to 1300 $\mu m$ through a constant
$L_{1300}/L_{60}$ luminosity ratio (thick line). The three estimates 
are consistent within the errors. 

A precise knowledge of the LLF allows a model-independent estimate of
the minimal contribution of galaxies to the IR-mm extragalactic
BKG. This minimal estimate, which is proportional to the local volume
emissivity $j_{\nu}=\int dL\ \rho(L)L$, was computed assuming no
evolution and that galaxies exist only up to z=1.
The minimal galaxy BKG is shown in Fig. 3 as the lower shaded
region: given such conservative assumptions, the real extragalactic
BKG is quite likely in excess of this.
Note that, because of the strong K-correction effect for even such low-z
objects, this spectrum is much
broader than the average local long-$\lambda$ spectral emissivity
of galaxies $j_{\nu}$.

\begin{figure}	
\vspace{130pt}
\caption{Contributions of galaxies to the extragalactic BKG (see text).
} 
\end{figure}

The upper shaded region in Fig. 3 is a recent estimate (Puget et
al. 1996) of the extragalactic BKG in the sub-mm cosmological window,
where the foreground emissions from the Galaxy and the interplanetary dust and
the Cosmic BKG Radiation are at the minimum. 
There is a
wide margin between this and our minimal BKG, which may indicate strong
evolution. 

In principle, the comparison between the local emissivity $j_{\nu}$
and the BKG spectrum allows inferences about the average redshift of 
the emitting sources. In practice, significant uncertainty is 
introduced by the spectral evolution with cosmic time. Fig. 3 reports two 
predictions corresponding to quite different models of galaxy formation and 
evolution: a merging scheme (with most of the flux contributed by $z<1$, 
dashed line) and a pure luminosity evolution model (with contributions at
$z>1$, continuous line). 
The two predict quite similar BKG's, the former assuming a constant
source spectrum, the latter one getting warmer in the past following the
higher star-formation rate and the increased average radiation field.

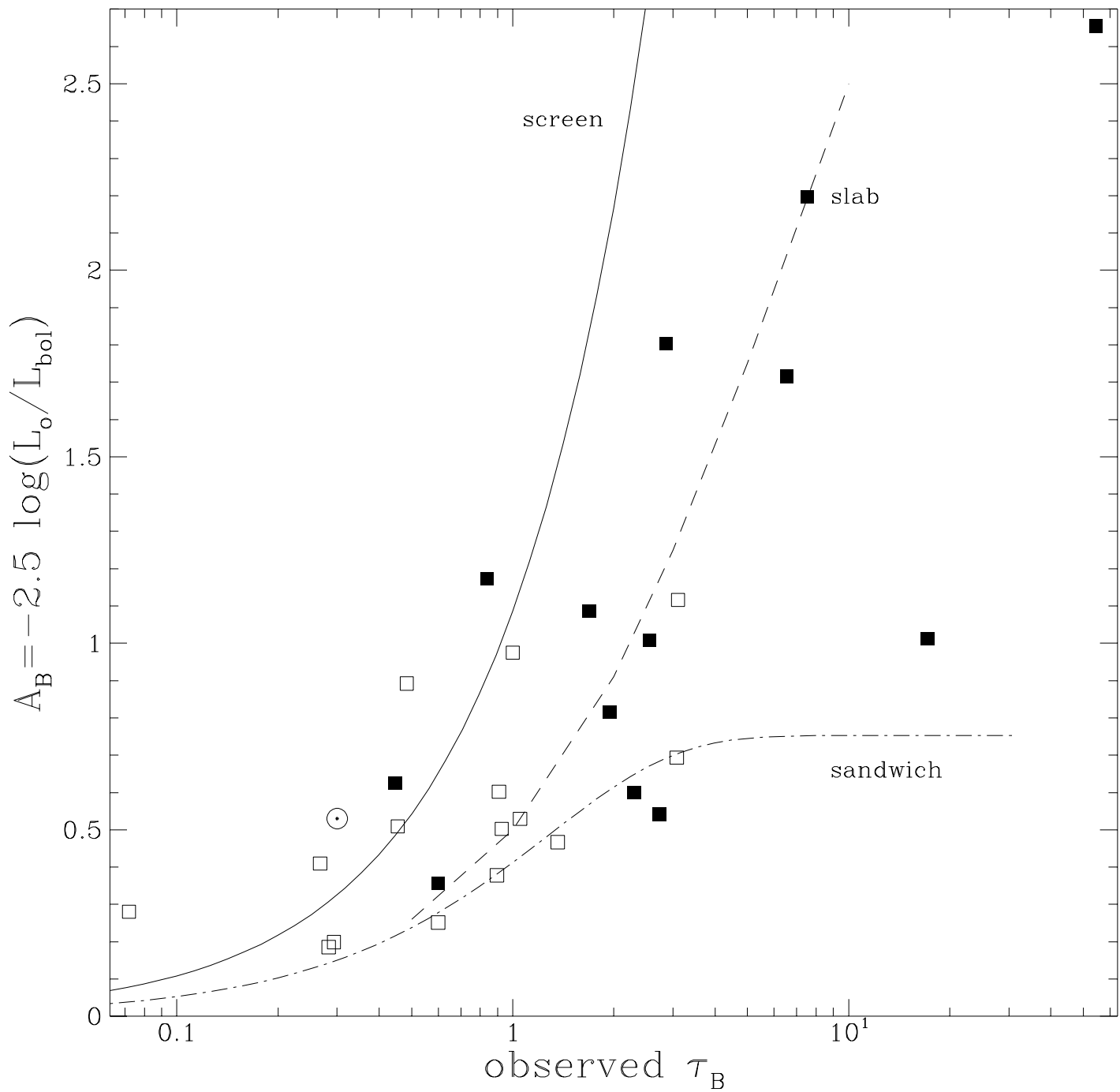
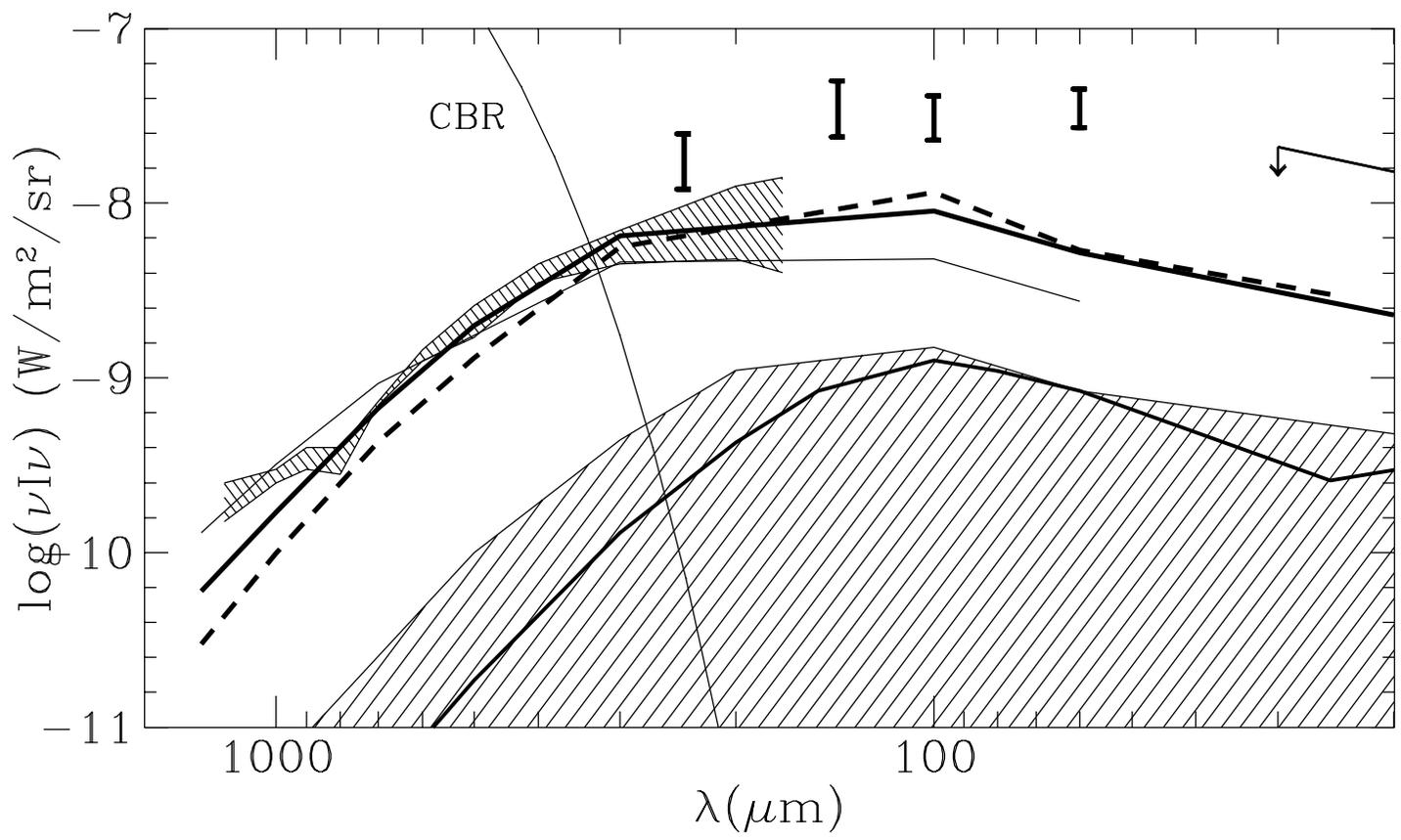
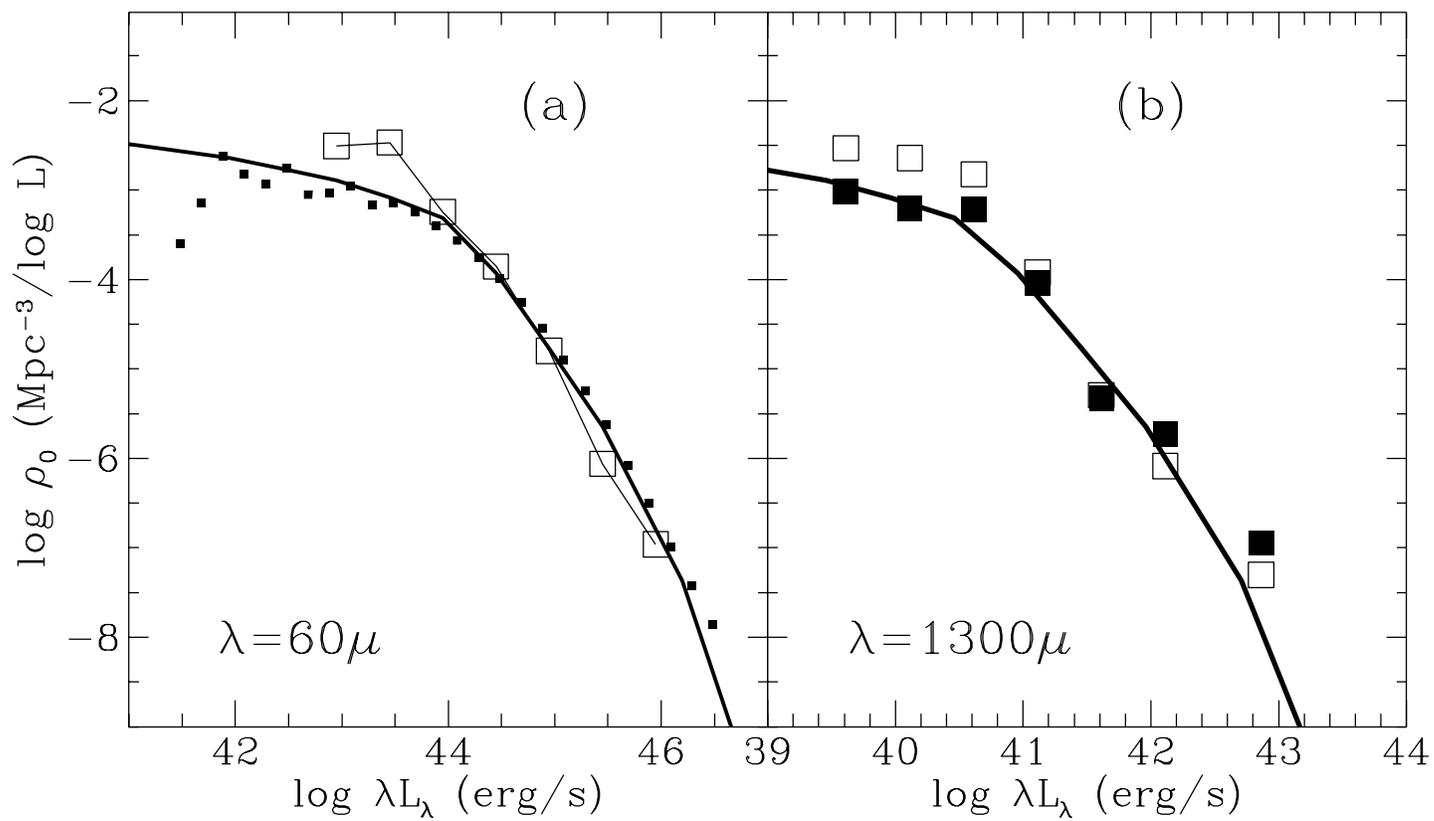
\end{document}